\documentclass[
    prb,
    twoside,
    twocolumn,
    10pt,
    floatfix,
    superscriptaddress,
]{revtex4-1}

\usepackage{glossaries}
\usepackage{graphicx}
\usepackage[version=4]{mhchem}
\usepackage{siunitx}
\usepackage[dvipsnames]{xcolor}

\usepackage[colorlinks=true]{hyperref}
\hypersetup{
    pdftoolbar=true,        
    pdfmenubar=true,        
    pdffitwindow=false,     
    pdfstartview={FitH},    
    pdftitle={A Computational Assessment of the Efficacy of Halides as Shape-Directing Agents in Nanoparticle Growth},
    pdfauthor={Joakim Löfgren, J. Magnus Rahm, Joakim Brorsson, and Paul Erhart},
    pdfnewwindow=true,      
    colorlinks=true,        
    linkcolor=NavyBlue,     
    citecolor=NavyBlue,     
    filecolor=NavyBlue,     
    urlcolor=NavyBlue       
}

\newcommand{\chalmersphys}{
    Chalmers University of Technology,
    Department of Physics,
    Gothenburg, Sweden
}
\newcommand{\chalmerschem}{
    Chalmers University of Technology,
    Department of Chemistry and Chemical Technology,
    Gothenburg, Sweden
}

\DeclareSIUnit\molar{\mole\per\cubic\deci\metre}
\DeclareSIUnit\Molar{M}
\DeclareSIUnit\atm{atm}
\DeclareSIUnit\monolayer{ML}

\newacronym{ctab}{CTAB}{cetyltrimethylammonium bromide}
\newacronym{cx}{CX}{van-der-Waals density functional with consistent-exchange}
\newacronym{dft}{DFT}{density functional theory}
\newacronym{gga}{GGA}{generalized gradient approximation}
\newacronym{lda}{LDA}{local density approximation}
\newacronym{mae}{MAE}{mean absolute error}
\newacronym{mare}{MARE}{mean absolute relative error}
\newacronym{mape}{MAPE}{mean absolute percentage error}
\newacronym{md}{MD}{molecular dynamics}
\newacronym{np}{NP}{nanoparticle}
\newacronym{pbe}{PBE}{Perdew-Burke-Ernzerhof}
\newacronym{pbesol}{PBEsol}{Perdew-Burke-Ernzerhof for solids}
\newacronym{rto}{RTO}{regular truncated octahedron}
\newacronym{si}{SI}{Supplementary Information}
\newacronym{ws}{WS}{Wulff shape}
\newacronym{xc}{XC}{exchange-correlation}

\newcommand{\eq}[1]{Eq.~\eqref{#1}}
\newcommand{\ion}[1]{\ce{#1^-}}
\newcommand{\angs}{\text{\normalfont\AA}}
\renewcommand{\vec}[1]{\ensuremath\boldsymbol{#1}}
\providecommand{\onlinecite}[1]{\hspace{-1 ex}\nocite{#1}\citenum{#1}}

\begin{document}

\title{
    A Computational Assessment of the Efficacy of Halides \texorpdfstring{\\}{}
    as Shape-Directing Agents in Nanoparticle Growth
}
\author{Joakim L\"{o}fgren}
\author{J. Magnus Rahm}
\affiliation{\chalmersphys}
\author{Joakim Brorsson}
\affiliation{\chalmerschem}
\author{Paul Erhart}
\email{erhart@chalmers.se}
\affiliation{\chalmersphys}

\begin{abstract}
We report a comprehensive study of aqueous halide adsorption on nanoparticles of gold and palladium that addresses several limitations hampering the use of atomistic modeling as a tool for understanding and improving wet-chemical synthesis and related applications.
A combination of thermodynamic modeling with density functional theory (DFT) calculations and experimental data is used to predict equilibrium shapes of halide-covered nanoparticles as a function of the chemical environment.
To ensure realistic and experimentally relevant results, we account for solvent effects and include a large set of vicinal surfaces, several adsorbate coverages as well as decahedral particles.
While the observed stabilization is not significant enough to result in thermodynamic stability of anisotropic shapes such as nanocubes, non-uniformity in the halide coverage indicates the possibility of obtaining such shapes as kinetic products.
With regard to technical challenges, we show that inclusion of surface-solvent interactions lead to qualitative changes in the predicted shape. Furthermore, accounting for non-local interactions on the functional level yields a more accurate description of surface systems.
\end{abstract}

\maketitle

\section{Introduction}

Wet-chemical synthesis has emerged as one of the primary routes for obtaining metal \glspl{np} with tailored properties.
One of the most important processes in the aqueous synthesis environment is the
adsorption of ions on the \gls{np} surface.
Detailed understanding of the influence of ion adlayers on growth, morphology, and physicochemical properties of \glspl{np} is therefore crucial for the continued development of nanotechnology.

Halides are one of the prototypical ionic species found in aqueous \gls{np} solutions as components of common metal salts, surfactants, and cleaning agents \cite{LohBurSca14, GhoMan18}.
In aqueous solution, halides tend to form adlayers on metal surfaces \cite{MagOckAdz95, Mag02}, resulting in lower surface energies and stabilization of higher-index facets.
The propensity of halides to adsorb is highly sensitive to the atomic structure of the surface, resulting in facet-dependent adlayer structures.
Since this can in turn stabilize different \gls{np} morphologies \cite{NikEl01, JohDujDav02, ChaOyaHir07} and guide \gls{np} growth, halides are commonly considered as ``shape-directing agents''.
For instance, during the growth of gold nanorods bromide plays an integral role in promoting the growth of, initially spherical, seed particles into rods \cite{HaKooChu07, SiLedDel12, HubTesSpa08, MeeCelSch16, AlmNovWhi14}.
Similarly, several studies have found the presence of bromide conducive to Pd nanocube growth \cite{ChaOyaHir07, LimJiaTao09, PenXiePar13}.
Halides can also influence \gls{np} growth through other pathways, including the formation of metal complexes with the precursor and surface smoothing through oxidative etching \cite{XioCheWil05, ZheZenRud14}.

While the importance of parameters such as halide species, concentration, and counter ion is generally acknowledged, a quantitative understanding of their impact has yet to be achieved.
This can partly be attributed to a gap between experimental and theoretical work.
Experiments take place in a complex chemical environment where the metal precursor commingles with other reactants such as surfactants and reducing agents, which is often neglected completely in modeling work.
This holds true in particular for \gls{dft} calculations, which despite being the main workhorse of computational surface science face technical challenges, e.g, with respect to obtaining accurate surface energies \cite{SinMar09, PatBatSun17} and accounting for solvation effects \cite{AstPraKre03, GroSchDra03, PerWalLaz12}.
A comprehensive analysis of halide adsorption must therefore address multiple aspects.
Here, as a step toward this goal, we present an extensive quantitative investigation of halide adsorption in aqueous solution on Au and Pd surfaces as well as \glspl{np}, which are of interest for a wide range of applications in nanomedicine \cite{Mur07,DumCou15,TraDeGPie17}, clean energy \cite{WadNugLid15,NamOguOhn18} and catalysis \cite{Tho07,Per12,BejGhoSar16}.
We employ a thermodynamic model that uses a combination of data from \gls{dft} calculations and experimental free energies to create Wulff constructions.
Our model enables quantitative predictions for equilibrium shape and surface area density of halide-covered \glspl{np} as a function of concentration in the range relevant for synthesis.
In contrast, previous \gls{dft}-based investigations into the thermodynamics of halide adsorption on metal surfaces have largely approached the problem from an electrochemical perspective, describing the stable adsorbate geometry on single surfaces as a function of the electrode potential \cite{RomGosFor14,GosRomGro15,GroSak19}.
The combination \gls{dft} calculations with experimental data allows us to compensate for short comings in \gls{dft} calculations, including ionization and solvation, without increasing the computational burden.
The inclusion of a large number of high-index facets renders the predicted shapes more realistic.
The next two sections introduce the key features of the thermodynamic model and provide a summary of the computational methodology, which is followed by results and discussion.
Additional material, including a detailed description of the modeling approach as well as complementary results, is presented in the \gls{si}.

\section{Thermodynamic model}\label{sect:model}

In equilibrium, a particle with fixed volume minimizes its surface free energy by assuming a shape described by a set of points
\begin{equation}
    \mathcal{W} =  \left\lbrace \boldsymbol{x}:\,\boldsymbol{x} \cdot \boldsymbol{n} \leq \gamma\left[\boldsymbol{n}\right]\, \text{for all}\, \boldsymbol{n} \right\rbrace,
    \label{eq:wulff}
\end{equation}
where $\gamma\left[\boldsymbol{n}\right]$ is the surface free energy \cite{Wul01}.
This is known as  Wulff's theorem and $\mathcal{W}$ is referred to either as the \gls{ws} or Wulff construction.
For a given crystalline \gls{np} of material \ce{M}, it is typically sufficient to consider only a small set of orientations described by Miller indices $\{hkl\}$.

If the twin boundary energy of $M$ is known, Wulff's theorem can be generalized to twinned particles, the most prominent examples are decahedra and icosahedra.
The latter comprise five and twenty tetrahedral grains, respectively, that are patched together to fill space \cite{Mar83, Mar84}.

We are interested in the case where a \gls{np} is in equilibrium with a source of adsorbates \ce{X}.
The free energy can then be written as the sum of the free energy of the clean surface and the change induced by adsorption of $N$ units of \ce{X}, i.e.
\begin{align}
    \gamma \left[\ce{M}(hkl):N\ce{X}\right]  = \gamma \left[\ce{M}(hkl) \right] + \Delta \gamma\left[\ce{M}(hkl):N\ce{X}\right].
    \label{eq:gamma}
\end{align}
For an aqueous halide $\ce{X^-_{aq}}$, we can approximate the surface free energy of adsorption on the right-hand side of \eq{eq:gamma} using total energies from \gls{dft} calculations by
\begin{equation}\label{eq:dgamma}
    \begin{split}
        \Delta & \gamma\left[\ce{M}(hkl):N\ce{X^-_{aq}}\right] \approx \frac{1}{A} \Bigl( E_{\text{DFT}}\left[ \ce{M}(hkl):N\ce{X} \right] \Bigr. -  \\
               & \Bigl. E_{\text{DFT}}\left[\ce{M}(hkl)\right] - N\left(\Phi\left[\ce{M}(hkl)\right] + \mu\left[\ce{X^-_{aq}}\right]\right) \Bigr)
    \end{split}.
\end{equation}
To avoid a charged slab calculation, a thermodynamic cycle \cite{SelPatOli96} is used to replace $E_{\text{DFT}}\left[\ce{M}(hkl):N\ce{X^-_{aq}}\right]$ that would otherwise appear in \eq{eq:dgamma} with the energy of a neutral system $E_{\text{DFT}}\left[\ce{M}(hkl):N\ce{X}\right]$ from which the appropriate multiple of the work function $\Phi\left[\ce{M}(hkl)\right]$ is subtracted.

In \eq{eq:dgamma}, the chemical environment, i.e., temperature, pressure, and concentration, is represented by the chemical potential $\mu$.
The surface free energy of adsorption is a linear function of $\mu$ where the slope is equal to the adsorbate surface area density $n_s = N/A$.
For fixed $(hkl)$, the slope is furthermore proportional to the coverage $\Theta = NA_{\text{prim}}/A$, where $A_{\text{prim}}$ is the area of the primitive surface cell.
To obtain the dependence of $\Delta \gamma$ on the chemical environment we assume ideal solution conditions, under which
\begin{align}\label{eq:mu}
    \mu\left[\ce{X_{aq}}\right]  = \mu^{\circ}\left[\ce{X_{aq}}\right] + k_{B}T \ln \left( \frac{c\left[\ce{X^-_{aq}}\right]}{c^{\circ}} \right).
\end{align}
Here, the $\circ$-superscript indicates that the quantity is given at standard state, defined by $P^\circ=\SI{1}{\bar}$ and $c^\circ=\SI{1}{\Molar}$.
The first term on the right-hand side of \eq{eq:mu} is thus the chemical potential at the standard state conditions.
Here, we calculate its value using a scheme in the spirit of Ref.~\onlinecite{PerWalLaz12}, where \gls{dft} calculations are combined with experimental reaction energies, in the present case taken from the curated and consistency-checked data provided in Ref.~\onlinecite{HunRei11}.
(Details are provided in the first section of the \gls{si}.)
Furthermore, the two first terms on the right-hand side of \eq{eq:dgamma} contain surface-solvent interactions.
Since explicit inclusion of water molecules at the \gls{dft} level is computationally expensive, they are accounted for here using an implicit solvation model \cite{MatSunLet14, MatHen16}.

\section{Computational details}\label{sect:comp}

All \gls{dft} calculations were carried out with the Vienna ab initio simulation package \cite{KreHaf93} (VASP, version 5.4.4), which uses plane-wave basis sets \cite{KreFur96} and the projector augmented wave method \cite{Blo94, KreJou99}.
To accurately model surfaces well as bulk systems, the \gls{cx} \cite{BerHyl14} was employed.
For comparison, a subset of the calculations was also carried out using the \gls{lda}, \gls{pbe},\cite{PerBurErn96} and \gls{pbesol} functionals \cite{PerRuzCso08}.
The plane-wave basis set was truncated at a cut-off energy of \SI{450}{\eV}, and the electronic levels were smeared using a Gaussian scheme with smearing parameter $\sigma = \SI{0.1}{\eV}$.
For Brillouin zone integration, Monkhorst-Pack and $\Gamma$-centered grids were used depending on the symmetry of the unit cell, with a $\vec{k}$-point density of $\SI{0.2}{\per\angstrom}$.
Geometry optimization was performed for all systems, during which the atomic positions were allowed to relax until the forces were less than \SI{10}{\meV\per\angstrom}.

For the calculations of the bare surface energies, we employed the extrapolation approach described in Refs.~\onlinecite{Boe94},\onlinecite{FioMet96} using $10$ slabs of different thickness for each $(hkl)$. For the adsorbate calculations, surfaces were modeled using slabs with an approximate thickness of $5a_0/\sqrt{3}$ where $a_0$ is the lattice constant of the parent crystal.
Spurious interactions with periodic images were minimized by introducing a vacuum region of \SI{20}{\angstrom} in the direction normal to the slabs.
To account for interactions between surface and solvent, the VASPsol implicit solvation model \cite{MatSunLet14, MatHen16} was applied to both clean and adsorbate-covered slabs.
In addition to the three low-index surfaces $(111)$, $(100)$, $(110)$,\footnote{For $(100)$ the missing-row reconstruction was included.} adsorption was also modeled on slabs of the high-index surfaces $(210)$, $(211)$, $(221)$, $(310)$, $(311)$, $(321)$, $(322)$, $(331)$, and $(332)$.
For each slab halide coverages corresponding to $\Theta \in \{1, 1/2, 1/3, 1/4\}\,\text{ML}$ were considered.
On the $(111)$ surface, the $\Theta = 1/3$ coverage was realized in a $\sqrt{3}\times\sqrt{3}$ unit cell.
For $(111)$, $(100)$, and $(110)$ we included all commonly considered high-symmetry sites.
For all other surfaces we employed Bayesian optimization to determine the most stable configurations \cite{TodGutCor19}.

The obtained surface energies were subsequently used to construct \glspl{ws} of single crystalline as well as decahedral and icosahedral \glspl{np} and average (facet-weighted) surface energies using the \textsc{wulffpack} Python package \cite{RahErh20}.

\section{Results and discussion}

\subsection{Clean surface properties}
We first assess the efficacy of various \gls{xc} functionals for accurately representing quantities important in the modeling of surface systems.
To this end, the lattice constant, clean surface energy as well as the work function are reported for the late non-magnetic face-centered cubic transition metals (\autoref{tbl:comp}).
The set of \gls{xc} functionals includes \gls{lda}, \gls{pbe}, and \gls{pbesol}, all of which are commonly used in surface calculations.
This set is extended by the non-local \gls{cx} functional, since several recent studies have highlighted the advantages of including non-local correlations for obtaining accurate results for surfaces \cite{LoeGroMot16, PatBatSun17} as well as bulk systems \cite{IlaZimWon15, GhaErhHyl17}.
For each combination of functional and quantity, \gls{mae} and \gls{mape} were calculated relative to available experimental data \cite{HaaTraBla09,Tys75,TysMil77,DerKerWor15}.

\begin{table}[htpb]
    \setlength{\tabcolsep}{4pt}
    \centering
    \caption{
        Comparison of different \gls{xc}-functionals with respect to properties relevant for surface systems.
        The following properties are included along with error estimations calculated with respect to available experimental data:
        lattice constants \cite{HaaTraBla09} (corrected for ZPE), surface energies \cite{Tys75, TysMil77}, and work functions \cite{DerKerWor15}.
        The calculated mean absolute errors show that the non-local \gls{cx} functional outperforms the other functionals included in the test, in particular for the clean surface energies.
    }
    \label{tbl:comp}
    \begin{tabular}{c c *{4}{c} c}
        \hline
        \noalign{\vskip 0.5mm}
        Quantity & Element & \multicolumn{4}{c}{XC Functional} & Exp. \\\cline{3-6}
        \noalign{\vskip 0.5mm}
                 &    & \gls{lda} & \gls{pbe} & \gls{pbesol} & \gls{cx} &  \\
                 \noalign{\vskip 2pt}
                 & Cu   &        $3.521$ &        $3.629$ &        $3.565$ &        $3.576$ &        $3.595$ \\
                 \noalign{\vskip 1pt}
                 & Ag   &        $4.002$ &        $4.147$ &        $4.051$ &        $4.060$ &        $4.063$ \\
                 \noalign{\vskip 1pt}
        $a_0$   & Au   &        $4.051$ &        $4.153$ &        $4.080$ &        $4.093$ &        $4.061$ \\
        \noalign{\vskip 1pt}
        (\angs) & Pd   &        $3.840$ &        $3.939$ &        $3.872$ &        $3.881$ &        $3.876$ \\
        \noalign{\vskip 1pt}
                & Pt   &        $3.897$ &        $3.967$ &        $3.916$ &        $3.930$ &        $3.913$ \\
                \noalign{\vskip 1pt}
                & MAE  &        $0.039$ &        $0.065$ &        $0.014$ &        $0.015$ &         \\
                \noalign{\vskip 1pt}
                & \textbf{MAPE} &  $\vec{1.0\%}$ &  $\vec{1.7\%}$ &  $\vec{0.4\%}$ &  $\vec{0.4\%}$ &   \\
                \noalign{\vskip 1pt}
                \hline
                \noalign{\vskip 1pt}
                & Cu   &       $1.88$ &        $1.36$ &        $1.71$ &       $1.86$ &       $1.79$ \\
                \noalign{\vskip 1pt}
                & Ag   &       $1.23$ &        $0.77$ &        $1.05$ &       $1.23$ &       $1.25$ \\
                \noalign{\vskip 1pt}
        $\bar{\gamma}$  & Au   &       $1.22$ &        $0.77$ &        $1.07$ &       $1.28$ &       $1.51$ \\
        \noalign{\vskip 1pt}
        (J$/\text{m}^2$) & Pd   &       $1.98$ &        $1.41$ &        $1.78$ &       $1.99$ &       $2.00$ \\
        \noalign{\vskip 1pt}
                         & Pt   &       $2.14$ &        $1.60$ &        $1.99$ &       $2.23$ &       $2.49$ \\
                         \noalign{\vskip 1pt}
                         & MAE  &       $0.15$ &        $0.63$ &        $0.29$ &       $0.12$ &        \\
                         \noalign{\vskip 1pt}
                         & \textbf{MAPE} &  $\vec{8\%}$ &  $\vec{35\%}$ &  $\vec{16\%}$ &  $\vec{6\%}$ &   \\
                         \noalign{\vskip 1pt}
                         \hline
                         & Cu   &       $5.19$ &       $4.82$ &       $4.95$ &       $4.99$ &       $4.90$ \\
                         \noalign{\vskip 1pt}
                         & Ag   &       $4.81$ &       $4.37$ &       $4.55$ &       $4.57$ &       $4.53$ \\
                         \noalign{\vskip 1pt}
        $\Phi(111)$    & Au   &       $5.49$ &       $5.18$ &       $5.26$ &       $5.34$ &       $5.33$ \\
        \noalign{\vskip 1pt}
        (eV)         & Pd   &       $5.64$ &       $5.27$ &       $5.35$ &       $5.42$ &       $5.67$ \\
        \noalign{\vskip 1pt}
                     & Pt   &       $6.06$ &       $5.71$ &       $5.82$ &       $5.89$ &       $5.91$ \\
                     \noalign{\vskip 1pt}
                     & MAE  &       $0.18$ &       $0.20$ &       $0.11$ &       $0.08$ &        \\
                     \noalign{\vskip 1pt}
                     & \textbf{MAPE} &  $\vec{4\%}$ &  $\vec{4\%}$ &  $\vec{2\%}$ &  $\vec{2\%}$ &   \\
    \end{tabular}
\end{table}

\begin{figure*}[htb]
    \centering
    \includegraphics{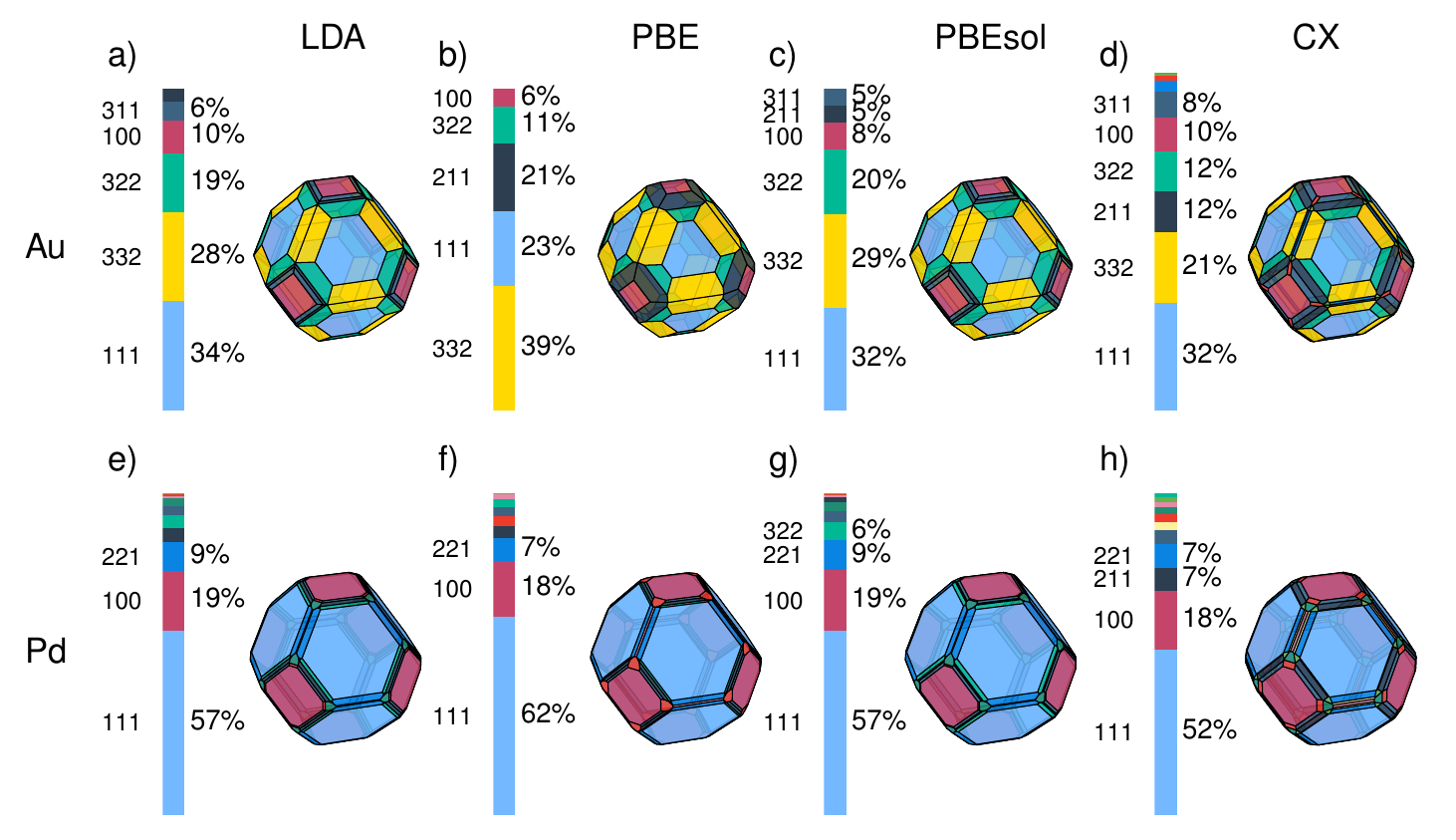}
    \caption{
        Wulff shapes for clean Au and Pd \glspl{np} in vacuum using surface energies obtained with different \gls{xc} functionals.
        The colored bars show the fractional area occupied by different facets.
        Changing the functional can lead to qualitative differences in the predicted shape as seen in the case of \gls{pbe} for Au where \{332\} facets represent the largest fraction in contrast to the other functionals.
        Despite giving quantitatively different results for the surface energies of individual facets, \gls{pbesol} and \gls{lda} predict similar shapes since the latter are determined by the surface energy \emph{ratios} rather than the absolute values.
    }
    \label{fig:wulff_clean}
\end{figure*}

For the lattice constants, \gls{pbesol} and \gls{cx} both offer comparable improvement over \gls{lda} and \gls{pbe}, respectively.
The calculated values are in quantitative agreement with previous benchmarks for bulk systems \cite{GhaErhHyl17}, where it was found that \gls{cx} outperforms \gls{pbesol} also when a wider range of bulk properties and materials is considered.

Experimentally, surface energies are commonly derived from surface tension measurements on liquid droplets, which provide a facet-weighted average surface energy \cite{TraXuRad16}.
Here, a significant improvement is seen in the \gls{mape} values in going from conventional \gls{gga} functionals to \gls{cx}.
In particular, while \gls{pbe} gives reasonable results for surface energies of simple metals \cite{TraXuRad16}, the \gls{mape} of 35\%{} found here suggests that caution should be exercised in the application of \gls{pbe} to late transition metals.
It is well-known that \gls{lda}, on the other hand, yields relatively accurate surface energies \cite{VitRubSkr98}, which can, however, be attributed to an error-cancellation effect \cite{PatBatSun17}.
As can be seen from the \gls{mape} values the systematic extension of \gls{dft} to include non-local interactions represented by \gls{cx} leads to a slight improvement over \gls{lda}.
The systematic increase in surface energy seen when comparing \gls{cx} to either \gls{pbe} or \gls{pbesol} reflects the contribution of non-local correlation to the energetic cost of cleaving a crystal to create a surface.

The $\{111\}$ work functions exhibit similar trends as the lattice constants, with \gls{pbe} and \gls{lda} systematically under and overestimating experimental data, respectively.
The \gls{mape} is reduced by about $2\%{}$ when either \gls{pbesol} or \gls{cx} are used instead, with the latter being slightly closer to the experimental values.

Overall, \gls{cx} outperforms the other functionals over the range of elements and properties included in this comparison.
Since both surface energy and work function appear explicitly in the surface free energy of adsorption in \eq{eq:dgamma}, \gls{cx} is thus the preferred functional for the present purpose.

While \gls{xc} functionals can exhibit large differences in predictions of surface energies, the equilibrium shape of a \gls{np} is sensitive to surface energy ratios rather than absolute values by virtue of Wulff's theorem \eq{eq:wulff}.
For this reason, we also compare actual \glspl{ws} obtained via the above set of functionals for Au and Pd.
To give an accurate representation of the \gls{ws}, the set of three low-index facets is further augmented by several high-index facets.

The \glspl{ws} for single crystalline particles (\autoref{fig:wulff_clean}) reveal that, while both Au and Pd particles have shapes derived from the \gls{rto} with additional faceting provided by high-index surfaces, the fractional area occupied by the facets varies not only in magnitude, but also in ordering when the \gls{xc} functional is changed.
The differences are most pronounced in the case of Au, where a large portion of the surface is occupied by high index facets.
For instance, \gls{pbe} predicts $\{332\}$ and $\{111\}$ as the facets with highest occupancy at $39\%{}$ and $23\%{}$, respectively.
With \gls{cx}, however, the order is reversed with $32\%{}$ of $\{111\}$ and only $21\%{}$ of $\{332\}$.
The\gls{pbe} results compare well with the Au and Pd Wulff constructions obtained using the same functional in Ref.~\onlinecite{TraXuRad16}.

In conclusion, the choice of \gls{xc} functional has ramifications that go beyond the degree of numerical agreement with experiment, and can lead to qualitative differences in predicted \gls{np} shapes.

\subsection{Particle morphology and the halogen group}
Before presenting any results related to halide adsorption, we note that typical wet-chemical synthesis concentrations are on the order of $0.01\text{--}\SI{1}{\Molar}$. Hence, we use $c\left[\ce{X^-_{aq}}\right] = \SI{0.1}{M}$ as
a representative value for aqueous states throughout the remainder of the text, unless the concentration dependence is explicitly considered.
Similarly, while variations in temperature and pressure can in principle be accounted for in \eq{eq:dgamma}, we restrict ourselves to ambient conditions.

\begin{figure}[htb]
    \centering
    \includegraphics{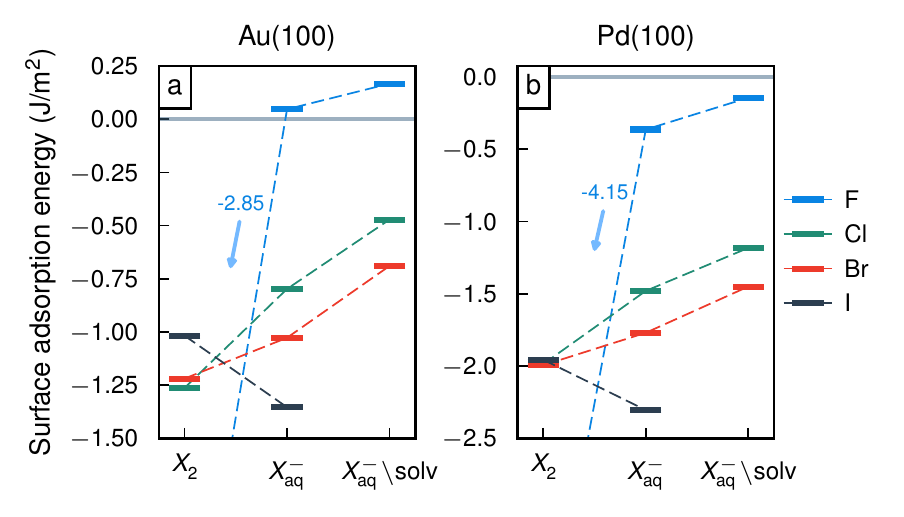}
    \caption{
        Adsorption energy level diagram elements in the halogen group on the $(100)$ surface of a) Au and b) Pd at $\Theta=\SI{1/2}{\monolayer}$.
        The halide concentration in the aqueous states is $\SI{0.1}{\Molar}$ corresponding to typical \gls{np} synthesis conditions.
        In going from dimerized state (\ce{X2}) to aqueous ion $(\ce{X^-_{aq}})$, the energetic ordering among the halogen species is reversed.
        To account for surface-solvent interactions an implicit solvation model can be used $(\ce{X^-_{aq}}\text{\textbackslash}\text{solv})$.
        The result is a non-negligible increase in adsorption energy comparable to the shift from dimer to aqueous ion.
    }
    \label{fig:halide_levels}
\end{figure}

The basic features of halogen adsorption on metal surfaces can be appreciated by observing the changes in surface adsorption energy as the adsorbate state is changed from the dimerized state to the aqueous ion.
This is illustrated in \autoref{fig:halide_levels} for the $(100)$ surface of Au and Pd.
The results for the aqueous ions are  given both with and without corrections for surface-solvent interactions provided by the VASPsol implicit solvation model.\footnote{
    Due to problems internal to VASPsol, we were not able to obtain data for the fully solvated state in the case of \ion{I} (\url{https://github.com/henniggroup/VASPsol/issues/34}; accessed: 2019/10/15).
}
Note that we use the thermodynamic sign convention where more negative surface free energies indicate greater stability. Furthermore, \autoref{fig:halide_levels} shows the surface adsorption energies as given by \eq{eq:dgamma} where the contribution from the clean surface is not included.
Negative (positive) energies thus correspond to exothermic (endothermic) reactions.
In the dimerized state, halogen adsorption strength correlates closely with electronegativity.
This is clearly observed, for example, on $\text{Au}(100)$ (\autoref{fig:halide_levels}a, left) where the most electronegative halogens are also more strongly adsorbed.
In going from dimer to solvated ion, however, the trend is reversed (\autoref{fig:halide_levels}a,b middle) and the adsorption strength decreases in the order $\ion{I} > \ion{Br} > \ion{Cl} > \ion{F}$.
This reversal reflects differences in ion-solvent interaction, where light halide ions interact more strongly with the solvent resulting in weaker adsorption \cite{HabBoc80}. Similar results have previously been obtained from PBE calculations complemented by thermodynamic data \cite{AlmNovWhi14}.
Including the surface-solvent interaction shifts the adsorption energy upward by more than \SI{0.2}{\eV}.
For \ion{Cl} and \ion{Br} on $\{100\}$, this shift is as large as the change in adsorption energy in going from dimerized state to aqueous ion, and for \ion{F} it can even result in a sign change of the adsorption energy.
Therefore, it is clear these interactions cannot, in general, be neglected.
\begin{figure*}[htb!]
    \centering
    \includegraphics{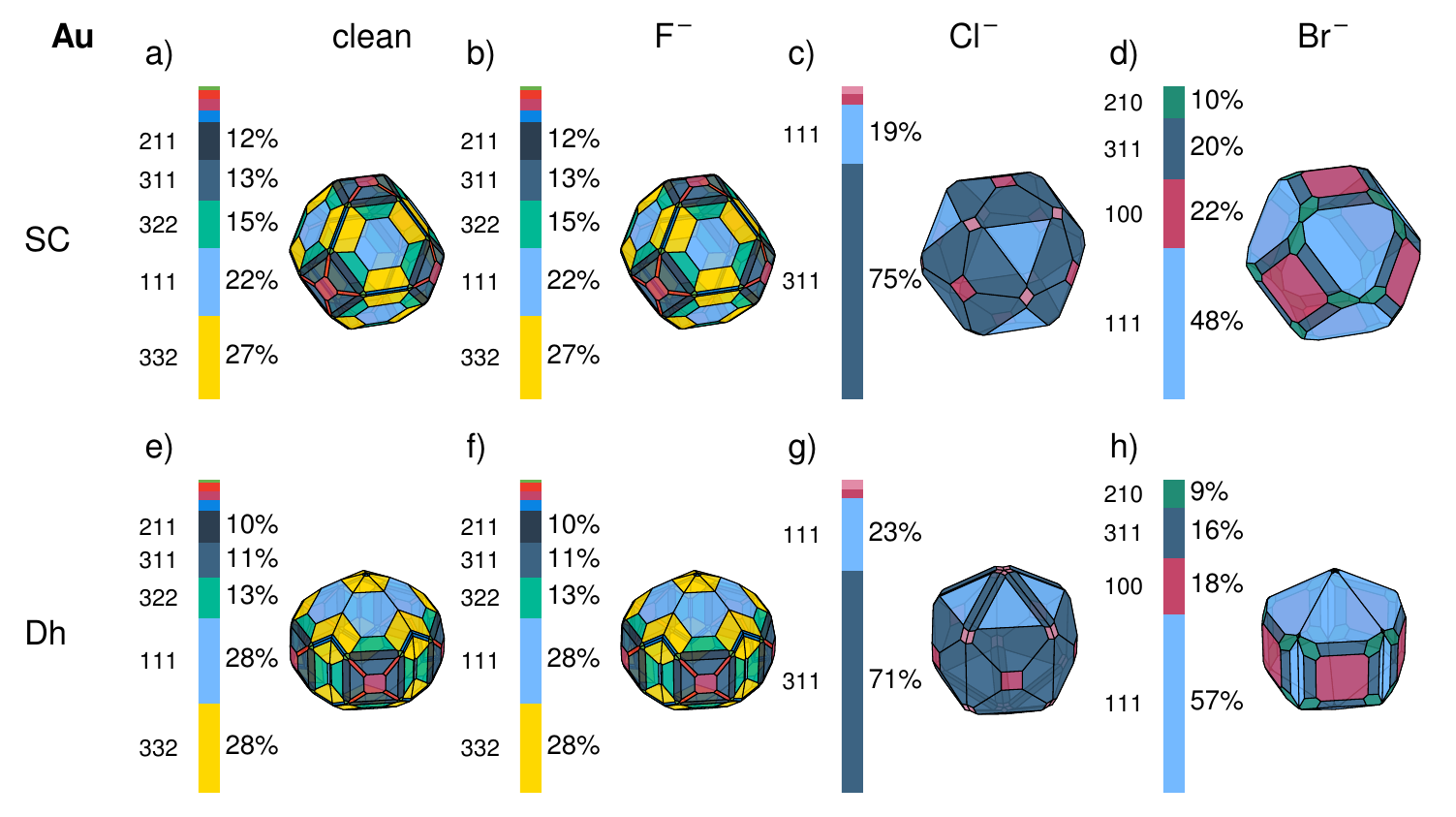}
    \caption{\label{fig:limits_Au}
        Wulff shapes predicted for single-crystalline (top) and decahedral (bottom) Au \glspl{np} immersed in $\SI{0.1}{\Molar}$ aqueous halide solutions under ambient conditions.
        Since \ion{F} adsorption is endothermic on Au it does not induce a shape change compared to the halide free case.
        Adsorption of \ion{Cl} is exothermic and results in the elimination of all vicinal facets except for $\{311\}$, while retaining some portion of \{111\} facets (c, g).
        \ion{Br}-covered \glspl{np}, on the other hand, feature a significantly increased amount of $\{111\}$ compared to the clean case and smaller amounts of $\{100\}$, $\{210\}$, and $\{310\}$ (d, h).
    }
\end{figure*}

We now consider the influence of halide adsorption on \gls{np} shape.
Here, the use of the generalized Wulff constructions enables us to also include decahedral \glspl{np} in our predictions.
This requires knowledge of twin boundary energies, for which we obtained
$\gamma_T\left[\ce{Au}\right] = \SI{23}{\milli\joule\per\meter\squared}$ and
$\gamma_T\left[\ce{Pd}\right] = \SI{68}{\milli\joule\per\meter\squared}$ with \gls{cx}, in good agreement with experimental estimates \cite{SuzBar58,XuLinFre91}.

On Au \glspl{np}, adsorption of \ion{F} (\autoref{fig:limits_Au}b,f) is largely endothermic, and a distribution of facets primarily comprising $\{111\}$ and vicinal facets (\autoref{fig:limits_Au}b, f) is obtained, identical to a clean \gls{np} under aqueous solvation conditions (\autoref{fig:limits_Au}a,e).
Adsorption of \ion{Cl} is, on the other hand, exothermic and leads to changes in faceting.
While the area fraction of $\{111\}$ remains similar to the clean case, with around $20\%{}$ for both single-crystalline and decahedral \glspl{np}, most high-index facets are eliminated.
Indeed, the facet distribution is dominated by $\{311\}$, which accounts for $70\%{}$ of the available area (\autoref{fig:limits_Au}c,g).
Upon replacement of \ion{Cl} by \ion{Br}, the area fractions again undergo significant changes resulting in a \gls{ws} with $50\%{}$ $\{111\}$, $20\%{}$ $\{100\}$ as well as smaller amounts of $\{210\}$ and $\{310\}$ (\autoref{fig:limits_Au}d,h).
\begin{figure*}[htb!]
    \centering
    \includegraphics{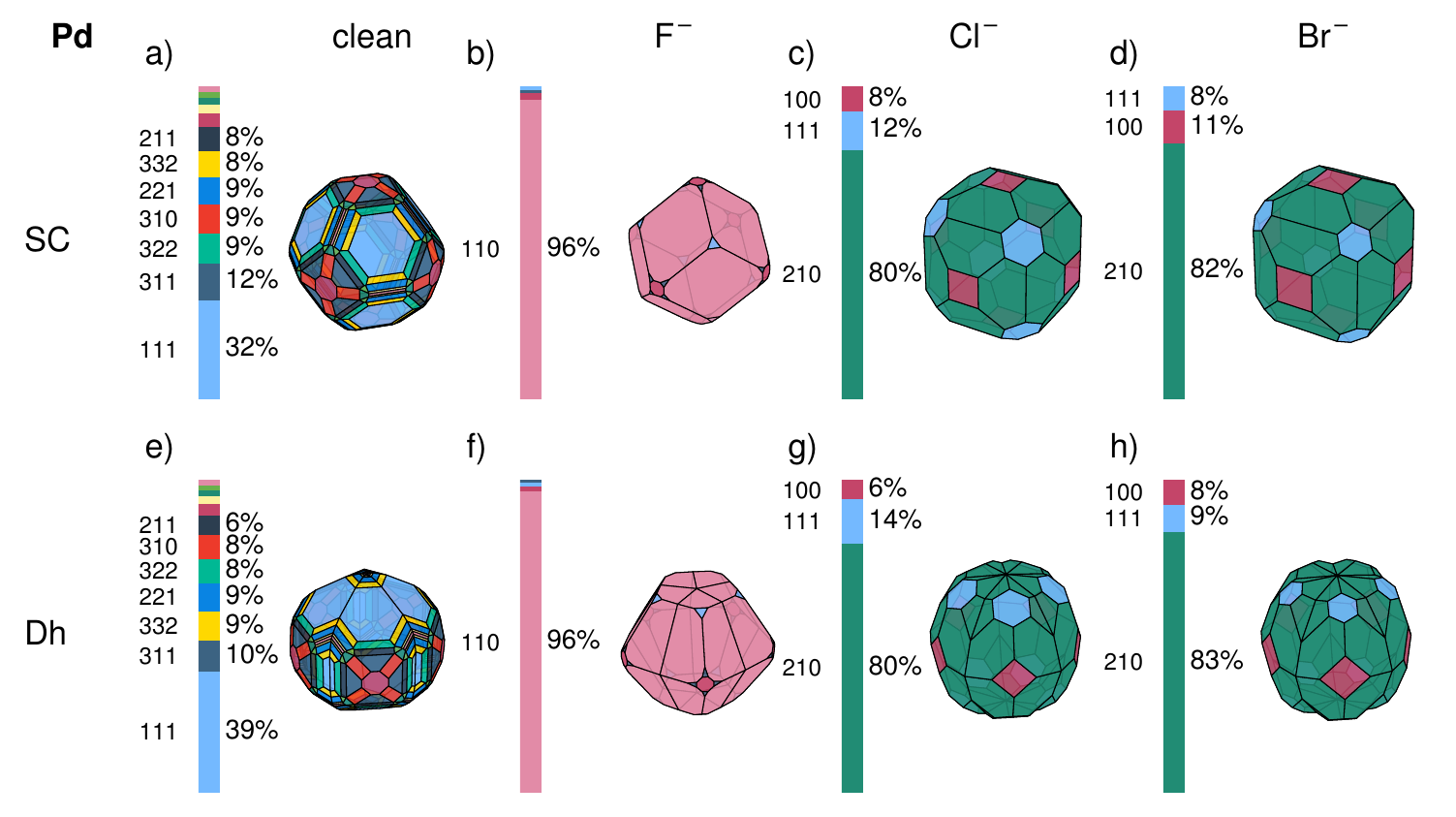}
    \caption{\label{fig:limits_Pd}
        Wulff shapes predicted for single-crystalline (top) and decahedral (bottom) Pd \glspl{np} immersed in $\SI{0.1}{\Molar}$ aqueous halide solutions under ambient conditions.
        In contrast to the case of Au, adsorption of \ion{F} on Pd \glspl{np} is exothermic and strongly favors $\{110\}$ facets.
        Consequently, single-crystalline (decahedral) \glspl{np} assume a rhombic dodecahedral (truncated pentagonal bipyramidal) shape.
        Pd \glspl{np} covered by \ion{Br} or \ion{Cl} adopt nearly identical shapes dominated by $\{210\}$ facets alongside minor amounts of \{111\} and \{100\}.
    }
\end{figure*}
In the case of Pd, \ion{F} adsorption can be exothermic, in particular $\{110\}$ facets are stabilized relative to clean, aqueous \glspl{np} (\autoref{fig:limits_Pd}a,b).
For single-crystalline \glspl{np} this yields a rhombic dodecahedron, while one obtains a truncated pentagonal bipyramid in the case of decahedral symmetry (\autoref{fig:limits_Pd}e,f).
In contrast to Au, the shapes obtained in the presence of \ion{Cl} or \ion{Br} are similar to each other (single-crystalline \glspl{np}: \autoref{fig:limits_Pd}c,d; decahedral \glspl{np}: \autoref{fig:limits_Pd}g,h) with about $80\%{}$ of their surfaces being $\{210\}$ facets alongside smaller fractions of $\{111\}$ and $\{100\}$.
We note that the prevalence of $\{210\}$ facets in this case is largely due to geometric constraints since the $(100)$ surface free energy is of similar magnitude for a wide range of concentrations (Fig.~S2).

In combination, the significant differences in faceting with halide adsorption and \gls{np} composition show that morphology prediction for halide-covered \glspl{np} is non-trivial.
It is also noteworthy that without the use of an implicit-solvation model, high-index facets do not appear in any of the above \glspl{ws} (Figs.~S3 and S4).
Hence, even if the adsorbate state is properly accounted for in the thermodynamic analysis, failure to include surface-solvent interactions can result in qualitatively incorrect predictions.
\begin{figure}[htb]
    \centering
    \includegraphics{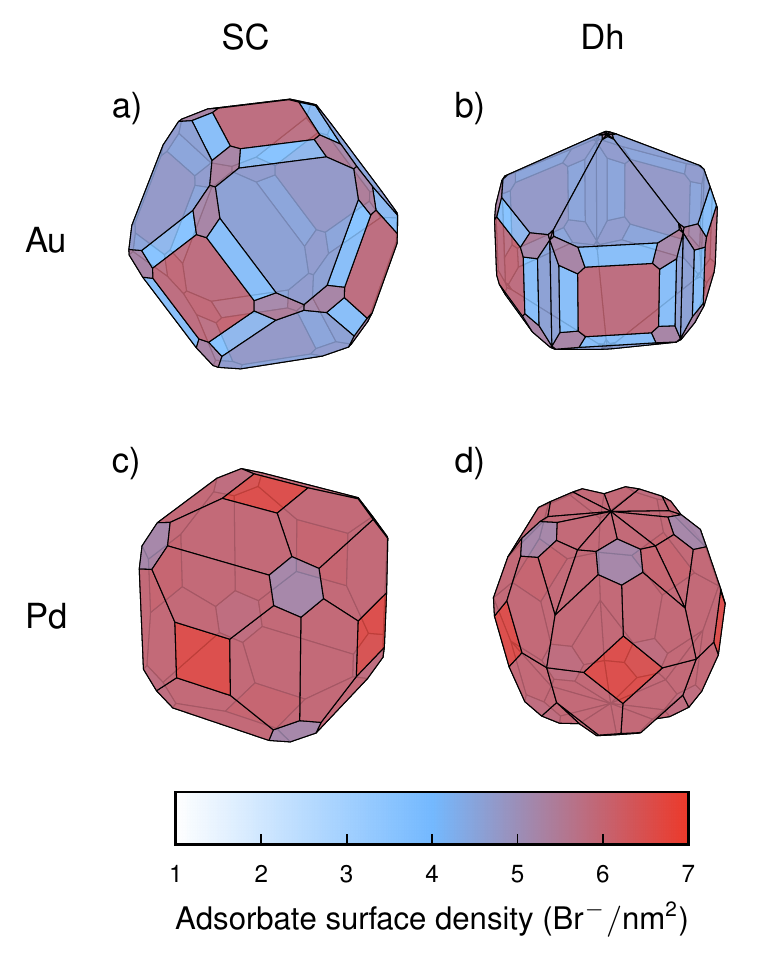}
    \caption{
        Bromide surface area density on \glspl{np} in a $\SI{0.1}{\milli\Molar}$ aqueous solution and ambient conditions.
        For both Au (a, b) and Pd (c, d) the surface area densities are non-uniform with the highest densities found on the $\{100\}$ facets.
        In the case of kinetically controlled growth, \ion{Br}-covered single crystalline (a, c) and decahedral \glspl{np} (b, d) can thereby template nanocubes or rods, respectively.
    }
    \label{fig:density}
\end{figure}
In addition to \gls{np} faceting, we can study the halide coverage, or surface area density.
The analysis leads to the important observation that for both Au and Pd \glspl{np}, the area density of adsorbates at fixed bulk concentration is facet dependent.
Considering for example \ion{Br} on Au (\autoref{fig:density}a) at a concentration of $\SI{0.1}{\Molar}$, the surface area density on $\{100\}$ and $\{111\}$ facets is $\SI{6.0}{\per\nm\squared}$ and $\SI{4.6}{\per\nm\squared}$, respectively, corresponding to coverages of $\SI{1/2}{\monolayer}$ and $\SI{1/3}{\monolayer}$.
This can yield anisotropic shapes since growth can be hampered in the case of high coverage.
In a kinetically controlled synthesis, in which the surface diffusion is slow compared to the arrival rate of monomers at the surface, this would promote growth on $\{111\}$ and turn octahedra into cubes \cite{QiBalZho15} while increasing the aspect ratio of decahedra \cite{XiaXieLiu13,QiCheYan19}.

\subsection{Sensitivity to the chemical environment}

Our thermodynamic model allows for the prediction of \gls{np} energetics and shape changes as a function of thermodynamic control parameters.
The most relevant parameter in this case is the halide concentration, which in a typical \gls{np} synthesis is on the order of \(10\text{--}\SI{1000}{\milli\Molar}\).
\begin{figure*}
    \centering
    \includegraphics{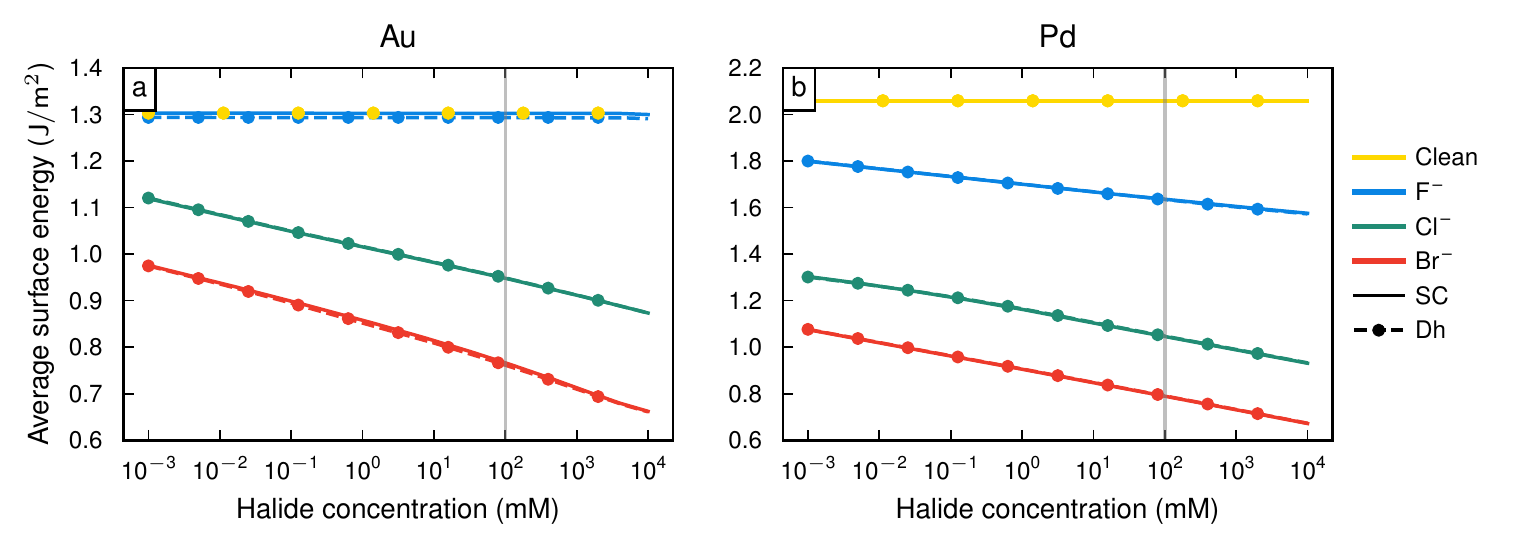}
    \caption{
        Weighted average surface energy of a \gls{np} gives a simple measure of the \glspl{np} stability against agglomeration or coalescence.
        The grey line indicates the concentration in \autoref{fig:halide_levels} to \ref{fig:limits_Pd}.
        The average surface energy decreases with increasing concentration and with increasing atomic number of the halide.
        For \ion{Br} concentrations on the order of $\SI{1}{\Molar}$, the average surface energy can decrease by as much as $50\%{}$ relative to that of clean, aqueous particles, thus suggesting the use of \ion{Br} as an effective capping agent.
    }
    \label{fig:poly_avg}
\end{figure*}
In applications where protecting the \glspl{np} against agglomeration is the primary concern, the reduction in the average (facet-weighted) surface energy $\bar{\gamma}$ with respect to the clean \gls{np} provides a measure of the overall stabilization that can be achieved by halide capping.
The stabilization follows the expected energetic ordering $\ion{Br}<\ion{Cl}< \ion{F}$ (\autoref{fig:poly_avg}; also see \autoref{fig:halide_levels}).
Increasing the halide concentration yields an exponential decrease in $\bar{\gamma}$ with the exception of \ion{F} on Au.
In the latter case, since the adsorption is endothermic on many gold facets, no stabilization is observed.
At a concentration of $\SI{0.1}{\Molar}$, \ion{Br} adsorption leads to a decrease of $\bar{\gamma}$ by $45\%{}$ and $60\%{}$ for Au and Pd, respectively.
Thus, halide adsorption can significantly lower the average surface energy of the \gls{np} providing a quantitative measure for the efficacy of halides as capping agents.
\begin{figure*}[htb!]
    \centering
    \includegraphics{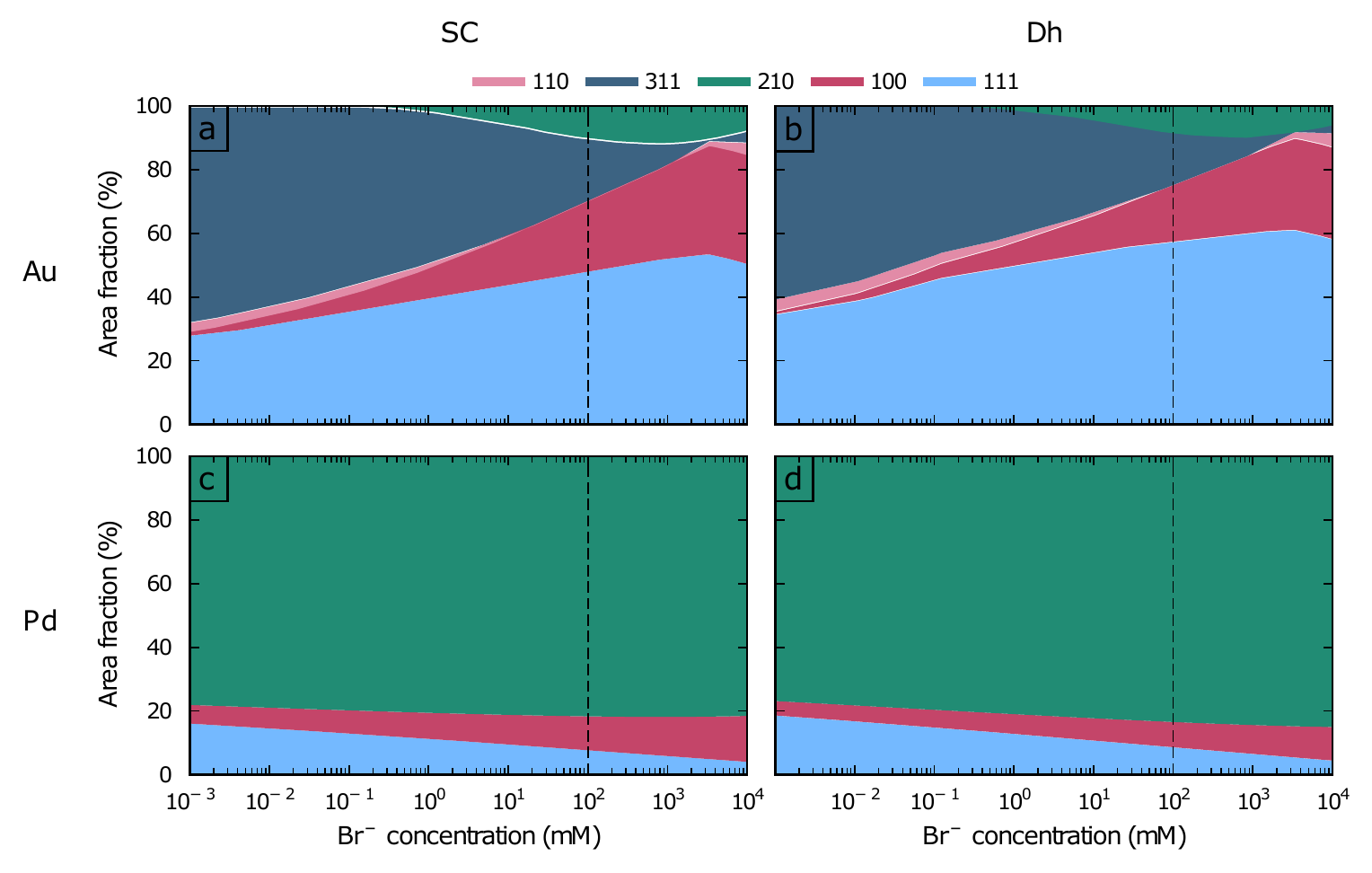}
    \caption{
        Evolution of \gls{np} faceting with increasing \ion{Br} concentration for single-crystalline and decahedral \glspl{np} of Au (top) and Pd (bottom).
        The dashed line indicates the concentration in \autoref{fig:halide_levels} to \ref{fig:limits_Pd}.
        The faceting of Au\glspl{np} is sensitive to the concentration, for instance the $\{310\}$ facets can almost be completely eliminated by increasing the concentration within the experimentally feasible range.
        For Pd\glspl{np}, the concentration dependence is weaker; an order of magnitude increase in concentration typically only results in a few percent increase of $\{100\}$ facets at the expense of $\{111\}$.
    }
    \label{fig:facet_conc}
\end{figure*}

To complement the \glspl{ws} presented in the previous section, we also map the facet area fractions as a function of the halide concentration (\autoref{fig:facet_conc}).
For brevity, the results presented here are restricted to \ion{Br}, since this is arguably the most important halide in shape-controlled \gls{np} synthesis.
The corresponding maps for \ion{F} and \ion{Cl} can be found in the \gls{si} (Figs.~S5 and S6.).

We observe that the sensitivity to Br concentration is largest for Au \glspl{np}, and crucially, that qualitative changes to the faceting can occur within the range of experimentally relevant concentrations.
This is most clearly illustrated by the \{310\} facets, which can be eliminated by increasing the concentration to $\SI{3.3}{\Molar}$.
For comparison, the area fraction at the representative concentration $\SI{0.1}{\Molar}$ is $20\%{}$ (\autoref{fig:facet_conc}a,b dashed lines).
Correspondingly, the amount of $\{111\}$ ($\{100\}$) increases from $48\%{}$ ($22\%{}$) to $53\%{}$ ($35\%{}$).
It should furthermore be noted that the concentration dependence is non-monotonic for the entire set of facets.
For instance, the area fraction of $\{111\}$ facets attains a maximum at $\SI{3.3}{\Molar}$.

For Pd \glspl{np}, the facet area fractions (\autoref{fig:facet_conc}c,d) change monotonically with concentration and at a lower rate than for Au.
Regardless of the concentration, $\{210\}$ facets represent about $80\%{}$ of the surface area and as the concentration increases the fraction of $\{100\}$ facets grows at the expense of $\{111\}$.

The above analysis shows, in conjunction with the results presented for the halide substitutions (\autoref{fig:limits_Au} and \ref{fig:limits_Pd}), that the shape of Au and Pd \glspl{np} can be tuned to some extent by changing either halide species or concentration.
The effect is, however, not significant enough to render anisotropic shapes such as cubes thermodynamically stable.

\section{Conclusions and outlook}

In this study, we have demonstrated how \gls{dft} calculations combined with an implicit-solvation model and experimental data can be used to formulate a thermodynamic model with the ability to predict the equilibrium shape and surface area density of halides on aqueous \glspl{np} under realistic conditions, including finite temperature, pressure and halide concentration.
Our contribution thus helps bridge the gap between experimental and theoretical studies of aqueous \glspl{np}.

For Au and Pd particles, halide species and concentration play an important role in determining \gls{np} faceting.
In particular for Au \glspl{np}, qualitative differences in faceting can be obtained by varying the concentration of \ion{Cl} and \ion{Br}, the two halides most commonly encountered in wet-chemical synthesis.
Here, the non-monotonic response of the distribution of facet area fractions to changes in concentration underscores the subtlety of \gls{np} shape prediction in complex environments and underscores the need for accurate atomistic simulations.
We emphasize, however, that adsorption of halides alone cannot stabilize anisotropic shapes such as cubes or rod-like particles.

Furthermore, our calculations reveal that the halide surface area density is facet dependent and for both Au and Pd \glspl{np} the highest density is found on the $\{100\}$ facets.
Our results are thus consistent with the picture of halides as capping-agents that can selectively passivate one or more facets and, given the right geometry and kinetic conditions, promote their growth \cite{XiaXieLiu13}.

In light of the above conclusions regarding the relation between halides and shape anisotropy, it is evident that further investigations into the synergistic effects between halides and their cations, most notably \gls{ctab}, are required.
Some steps in this direction have already been taken using classical \gls{md} \cite{MeeSul13, MeeCelSch16} as well as \gls{dft} simulations \cite{AlmNovWhi14, JaiTehShe14}.
In the case of classical \gls{md}, the reliability of any force-field description of the halide-metal interface must be carefully evaluated since traditional force-fields do not account for effects such as charge transfer and polarization at the interface \cite{SenHonIsl16, GeaRamJam18}.
In the case \gls{dft} calculations, the thermodynamic model utilized in the present work is in principle straightforward to apply to systems where both anions and cations are present, but the accuracy is limited by the quality of the available thermodynamic data.

Another aspect of \gls{np} shape prediction that was deemed to be outside the scope of the present study concerns the treatment of finite size effects.
It is well-known that the Wulff construction does not account for such effects and the predicted \glspl{ws} can thus only be considered realistic for large particles.
Quantifying the limit for when a \gls{np} can be considered large in this sense is important given the high area fractions predicted here for high-index facets such as $\{210\}$ and $\{310\}$.
More precisely, from a geometric point of view a \gls{np} needs to be large enough to accommodate stepped surfaces in the first place, and the energetic penalty incurred by the facet edges and corners needs to be considered.
As a result, a small \gls{np} may end up exposing smaller amounts of vicinal facets then what is predicted by a particular \gls{ws}.

\section*{Acknowledgments}

This work was funded by the Knut and Alice Wallenberg Foundation (2014.0226), the Area of Advance Nano at Chalmers, the Swedish Research Council (2015-04153, 2018-06482), and the Strategic Swedish Research Initiative (RMA15-0052).
Computer time allocations by the Swedish National Infrastructure for Computing at C3SE (Gothenburg), NSC (Link\"oping), and PDC (Stockholm) are gratefully acknowledged.

\end{document}